\documentclass{iopjournal}
\usepackage{amsmath}
\usepackage{siunitx}

\renewcommand{\vec}[1]{\ensuremath{\boldsymbol{#1}}}

\renewcommand{\articletype}[1]{{\vspace*{-8mm}\noindent \Large \sf Preprint}
	\vspace*{8mm} \\ \noindent
	{\scriptsize \sf{\bfseries \MakeUppercase{#1}}}}
\begin{document}

\articletype{Paper} 

\title{Shaper-based dispersion scan for the characterization of polarization-shaped laser fields}

\author{D. Köhnke$^1$\orcid{0000-0003-1151-8625}, J.-E. Havekost$^1$\orcid{0009-0003-2040-2920}, N. R. Rother$^1$\orcid{0009-0003-8469-5558}, C. Kuntze$^1$\orcid{0009-0009-8452-4998}, L. Englert$^1$\orcid{0000-0002-2275-9136}, T. Bayer$^1$\orcid{0009-0005-2280-5672} and M. Wollenhaupt$^{1,*}$\orcid{0000-0002-0839-1494}}

\affil{$^1$Carl von Ossietzky Universit\"at Oldenburg, Institut f\"ur Physik, Carl-von-Ossietzky-Stra\ss e 9-11, D-26129 Oldenburg, Germany}

\affil{$^*$Author to whom any correspondence should be addressed.}

\email{matthias.wollenhaupt@uol.de}


\begin{abstract}
We present a pulse-shaper-based dispersion-scan (d-scan) framework for the combined generation and characterization of polarization-tailored femtosecond laser fields.
By integrating a programmable $4f$ pulse shaper with polarization-resolved d-scan measurements, the framework enables the programmable synthesis and reconstruction of complex time-dependent polarization states.
We demonstrate its capabilities for several classes of vector fields, including counter-rotating circularly polarized pulse pairs, oppositely chirped counter-rotating circularly polarized pulses, polarization-gate pulses, and multi-pulse sequences utilizing polynomial, periodic and discrete spectral phase functions.
The reconstructed temporal electric fields and time-dependent ellipticities show excellent agreement with the simulated target fields and accurately reproduce the defining features of each pulse class. 
By combining the versatility of programmable pulse shaping with the robustness of the d-scan technique, this approach provides a flexible platform for the generation and validation of polarization-tailored ultrashort laser fields in ultrafast photonics and light-matter interaction studies.	
\end{abstract}

\section{Introduction}
\label{sec:Introduction}
Ultrashort laser pulses with time-dependent polarization have emerged as powerful tools in ultrafast optics and photonics, enabling tailored interactions with matter across a wide range of applications:
Time-dependent polarization states have been exploited to manipulate atomic \cite{Eckart:2016:PRL:133202,Mancuso:2016:PRA:053406,Pengel:2017:PRL:053003,Kerbstadt:2019:NC:658} and molecular \cite{Brixner:2004:PRL:208301,Suzuki:2004:PRL:133005} ionization dynamics, control chiral light-matter interactions \cite{Boewering:2001:PRL:1187,Garcia:2003:JCP:8781,Lux:2012:ACIE:5001}, and induce molecular orientation \cite{Villeneuve:2000:PRL:542,Stapelfeldt:2003:RMP:543,Karras:2015:PRL:103001}.
They have also been used to steer atomic multi-photon excitation \cite{Dudovich:2004:PRL:103003,Koehnke:2023:NJP:123025}, spin dynamics \cite{Sokell:2000:JPB:2005,Kimel:2005:Nature:655,Stanciu:2007:PRL:047601,Bayer:2019:NJP:033001}, and high harmonic generation \cite{Sansone:2006:Science:443,Fleischer:2014:NP:543,Kfir:2015:NP:99}. 
Three-dimensional locally chiral light fields \cite{Ayuso:2019:NP:866}, generated by the non-collinear superposition of bichromatic fields, have been used to control multiphoton ionization \cite{Koehnke:2026:PRR:033048}. Full utilization of these capabilities requires not only the ability to synthesize pulses with complex, time-varying polarization profiles, but also reliable methods for their temporal characterization \cite{Walmsley:2009:AOP:308,Monmayrant:2010:JPB:103001,Trebino:2000}.\\
Over the past decades, significant progress has been achieved in both pulse shaping and pulse characterization. 
Fourier-transform pulse shaping \cite{Weiner:2011:OC:3669} has become an established and versatile approach to control the spectral amplitude, phase, and polarization profile of femtosecond-laser pulses.
In such setups, individual spectral components are manipulated using spatial light modulators, including liquid-crystal devices \cite{Weiner:1990:OL:326,Brixner:2000:APB:S119,Wefers:1993:OL:2032}, deformable mirrors \cite{Zeek:1999:OL:493}, micro-mirror arrays \cite{Hacker:2003:APB:711}, acusto-optic modulators \cite{Hillegas:1994:OL:737,Dugan:1997:JOSAB:2348}, and, more recently, metasurfaces \cite{Divitt:2019:Science:890}.
These techniques allow the synthesis of ultrashort pulses with almost arbitrary time-dependent polarization states.\\
In parallel, a range of pulse characterization techniques has been developed, including temporal methods based on correlation measurements \cite{Armstrong:1967:APL:16,Mindl:1983:APB:201}, spectral methods such as spectral interferometry (SI) \cite{Lepetit:1995:JOSAB:2467} and spectral phase interferometry for direct electric-field reconstruction (SPIDER) \cite{Iaconis:1998:OL:792}, and spectrogram-based (spectro-temporal) methods, the most prominent and widely used technique being frequency-resolved optical gating (FROG) \cite{Trebino:1993:JOSAA:1101}.
These established techniques are reviewed in textbooks \cite{Diels:2006,Rulliere:2005,Weiner:2009:1,Trebino:2000}, book chapters \cite{Wollenhaupt:2012:1047} and review articles \cite{Walmsley:2009:AOP:308,Monmayrant:2010:JPB:103001}.
They were originally developed to characterize linearly polarized pulses.
Early extensions to the characterization of polarization-shaped pulses included dual-channel SI, also termed POLLIWOG \cite{Walecki:1997:OL:81,Brixner:2002:APB:s133}, and tomographic ultrafast retrieval of transverse light E-fields (TURTLE) \cite{Schlup:2008:OL:267,Xu:2009:JOSAB:2363}. 
More recent developments include V-FROG \cite{IlanHaham:2021:JPP:034017}, time-domain ptychography \cite{Schweizer:2024:OE:24346} and the amplitude swing \cite{Barbero:2024:OE:10862} technique, 
underscoring the interest in the characterization of polarization-shaped femtosecond pulses and the growing demand for reliable characterization methods.\\
The employment of a pulse shaper for the characterization of ultrashort laser pulses offers multiple advantages \cite{Monmayrant:2010:JPB:103001}. 
By programming the applied spectral modulation, a single setup can mimic various characterization techniques without modification of the optical setup. 
Pulse sequences produced by a pulse shaper instead of an interferometer benefit from enhanced mechanical stability, due to the common-path design, and control of the pulse-to-pulse time-delay has been demonstrated with extreme precision \cite{Koehler:2011:OE:11638}.
Combining pulse shaping and characterization within the same optical setup reduces experimental complexity. 
Moreover, the non-linear element required by most diagnostic techniques can be placed directly at the experimental interaction region, enabling \textit{in situ} characterization \cite{vonVacano:2006:OL:1154}.
Shaper-assisted implementations of established characterization techniques include SAC-SPIDER \cite{vonVacano:2006:OL:1154,vonVacano:2007:JOSAB:1091}, FROG \cite{Galler:2008:APB:427} and correlation measurements \cite{Galler:2008:APB:427,Koehler:2011:OE:11638,Kerbstadt:2017:OE:12518}.
In addition, pulse shapers have enabled new characterization strategies such as multiphoton intrapulse interference phase scan (MIIPS) \cite{Lozovoy:2004:OL:775}.
MIIPS applies a periodic spectral phase to the test pulse and records a nonlinear optical signal as a function of a phase parameter.
The unknown spectral phase can then be directly retrieved from the measured MIIPS trace up to constant and linear phase terms.
Recently, building on a MIIPS variant \cite{Lozovoy:2008:OE:592}, a related pulse-characterization technique using an additional polynomial phase was introduced \cite{Miranda:2012:OE:688}. 
This technique, known as dispersion-scan (d-scan), scans the applied dispersion and records a nonlinear signal without requiring a reference pulse or interferometric stability, combining simplicity and robustness in its implementation.
The d-scan technique has recently been extended to the characterization of ultrashort pulses with time-dependent polarization states \cite{DiazRivas:2024:JPP:015003,Perez-Benito:2024:OLT:111273}.\\
Despite these advances, polarization pulse shaping and characterization are typically implemented on separate experimental platforms.
This separation limits flexibility, increases alignment complexity, and restricts applicability in settings requiring rapid or adaptive control of polarization states.
Experiments requiring active adjustment of the polarization state \cite{Brixner:2004:PRL:208301,Suzuki:2004:PRL:133005}, such as the adaptive control of high harmonics generation, would benefit from an integrated approach.\\
Here, we introduce a shaper-based d-scan framework that combines the generation and characterization of shaped ultrashort laser pulses with tailored temporal polarization profiles.
The programmable pulse shaper further enables the application of pure second-order dispersion without the higher-order dispersion typically introduced by wedge-based implementations \cite{Miranda:2012:OE:18732}.
We demonstrate the versatility of this concept by synthesizing and characterizing four classes of polarization-tailored pulses with direct relevance for ultrafast light-matter interactions:
\begin{itemize}
	\item[(i)] Counter-rotating circularly polarized (CRCP) pulses, routinely applied for photoelectron interference and the generation of free electron vortices \cite{NgokoDjiokap:2015:PRL:113004,Pengel:2017:PRL:053003,Kerbstadt:2019:NC:658},
	\item[(ii)] Oppositely chirped CRCP (OC-CRCP) pulses, used for the creation of shaped free electron vortices \cite{Strandquist:2022:PRA:043110,Koehnke:2024:PRA:053109} and enabling non-perturbative excitation via rapid adiabatic passage in a V-type system \cite{Koehnke:2025:PRA:023104},
	\item[(iii)] Polarization-gate (PG) pulses, introduced for the generation of isolated attosecond pulses \cite{Corkum:1994:OL:1870,Sansone:2006:Science:443},
	\item[(iv)] Polarization-tailored pulse sequences, enabling advanced pump-probe experiments, in particular PG pulses, with potential application in XUV interferometry and double-slit experiments \cite{Kaneyasu:2023:SR:6142}.
\end{itemize}
\section{Setup and methods}
\label{sec:exp_methods}
\begin{figure*}[htpb]
	\includegraphics[width=\textwidth]{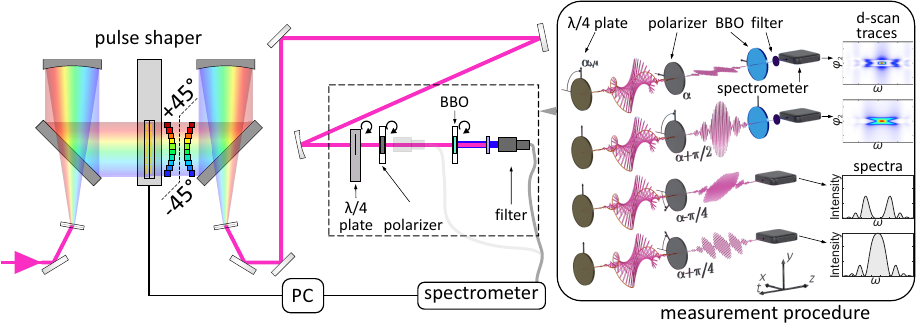}
	\caption{Setup for shaper based d-scan. Left: schematic $4f$ setup with shaper, $\lambda/4$-wave plate, polarizer, BBO and spectrometer. Right: Visualization of the measurement procedure. A polarization-tailored laser pulse is projected onto orthogonal linear polarization states and shaper-based d-scans are performed. Spectra for two additional projections are recorded to determine the relative constant and linear phase between the retrieved projections. \label{fig1}}
\end{figure*}

The experimental setup for the generation of polarization-tailored laser pulses and their characterization using the d-scan technique is shown in Fig.~\ref{fig1}.
The generation of polarization-shaped pulses using our home-built pulse shaper has been described in detail in \cite{Kerbstadt:2017:OE:12518,Kerbstadt:2017:JMO:1010}. 
Here, we briefly recapitulate in Section~\ref{subsec:PulseGeneration} the notation for the pulse generation, while Sec.~\ref{subsec:PulseCharacterization} describes the shaper-based characterization of vectorial pulses, building on the approach recently introduced in \cite{DiazRivas:2024:JPP:015003}.
\subsection{Generation of polarization-tailored laser pulses}
\label{subsec:PulseGeneration}
Near-infrared laser pulses with a duration of \SI{25}{\femto\second}, a central wavelength of \SI{790}{\nano\meter}, and a pulse energy of \SI{0.9}{\milli\joule} are spectrally broadened in a neon-filled hollow-core fiber.
By variation of the gas pressure between \SI{0}{\bar} to \SI{1}{\bar}, spectra supporting a transform-limited pulse duration of approximately \SI{7}{\femto\second} to \SI{9}{\femto\second} are generated.
The spectrally-broadened pulses are subsequently compressed and phase-modulated using a home-built $4f$ pulse shaper.
Prior to polarization shaping, the residual spectral phase of the input pulse is retrieved using the d-scan technique.
A representative d-scan trace and corresponding temporal intensity profile after compression are shown in Fig.~\ref{fig2} and discussed in Sec.~\ref{subsec:compcharac}.\\
Polarization shaping is implemented using a 640-pixel double-layer liquid-crystal spatial light modulator, which enables independent phase modulation of two orthogonal polarization components.
The temporal polarization profile is controlled by applying spectral phase functions $\varphi^{A/B}(\omega)$ to the two modulator layers A and B, which will be described below.
A schematic illustration is provided in Fig.~\ref{fig1} (left panel).
An achromatic $\lambda/4$ wave plate (\textsc{Bernhard Halle Nachfl.}) is used to convert the resulting field into circularly polarized light. 
Oriented at $\alpha_{\lambda/4}=\SI{0}{\degree}$, the wave plate converts the orthogonal linearly polarized components into circularly polarized components with opposite handedness, whereas for $\alpha_{\lambda/4}=\SI{45}{\degree}$ no polarization conversion occurs.
The total positive frequency spectrum is given as 
\begin{equation}
	\vec{\tilde{E}}^+(\omega)=\boldsymbol{\mathsf{\Lambda}}_{\lambda/4}(\alpha_{\lambda/4})\tilde{\mathcal{E}}(\omega-\omega_0)\left(e^{-i\varphi^{A}(\omega)}\vec{e}_{-\pi/4}+e^{-i\varphi^{B}(\omega)}\vec{e}_{+\pi/4}\right),
\end{equation}
where $\boldsymbol{\mathsf{\Lambda}}_{\lambda/4}(\alpha_{\lambda/4})$ denotes the Jones matrix of the $\lambda$/4 wave plate at an angle $\alpha_{\lambda/4}$, $\tilde{\mathcal{E}}(\omega)$ is the spectrum of the temporal envelope $\mathcal{E}(t)$, and $\omega_0$ is the central frequency.
The unit vectors $\vec{e}_{\pm\pi/4}=\frac{1}{\sqrt{2}}(1,\pm1)$ correspond to linear polarization along the optical axes of the modulator at $\SI{\pm45}{\degree}$.
The spectral phases applied to the two polarization components are parameterized as
\begin{align}
	\varphi^{A/B}(\omega) &= \phi^{A/B}_0+\phi^{A/B}_1(\omega-\omega_\mathrm{mod})+\phi^{A/B}_2(\omega-\omega_\mathrm{mod})^2\notag\\ &+A^{A/B}_\mathrm{sin}\sin\left[\tau^{A/B}_\mathrm{sin}(\omega-\omega_\mathrm{mod})+\phi^{A/B}_\mathrm{sin}\right]+\varphi_2(\omega-\omega_\mathrm{mod})^2,
	\label{eq:Phaseparameterization}
\end{align}
where $\omega_\mathrm{mod}$ is the modulation frequency, $\phi_0$, $\phi_1$ and $\phi_2$ are the constant, linear and quadratic phase coefficients.
$A_\mathrm{sin}$, $\tau_\mathrm{sin}$, and $\phi_\mathrm{sin}$ denote the amplitude, period, and phase of the sinusoidal modulation, respectively.
The coefficient $\varphi_2$ corresponds to the second-order dispersion applied to record the d-scan traces.
\subsection{Characterization of polarization-tailored laser pulses}
\label{subsec:PulseCharacterization}
The d-scan setup and measurement procedure for reconstructing the vectorial temporal electric field are illustrated schematically in Fig.~\ref{fig1} (center and right panels).
The field is projected onto the transmission axis of a broadband polarizer (\textsc{codixx AG}) oriented at an angle $\alpha$.
A \SI{5}{\micro\meter}-thick $\beta$-barium borate (BBO) crystal (\textsc{EKSMA optics}), with its optical axis aligned along the projection direction, is employed for second-harmonic generation.
The generated signal is separated from the fundamental using spectral filters (\textsc{Thorlabs FGB37-A} and \textsc{Thorlabs FD1B}).
D-scan traces are recorded for two orthogonal polarization projections at angles $\alpha$ and $\alpha+\pi/2$.
These measurements allow reconstruction of the vectorial spectrum up to an unknown constant phase offset $\Delta\varphi$ and linear phase term $\omega\tau$, according to 
\begin{equation}
	\vec{\tilde{E}}^+(\omega;\Delta\varphi,\tau)=\tilde{E}^+_\mathrm{\alpha}(\omega)\vec{e}_\alpha+\tilde{E}^+_\mathrm{\alpha+\frac{\pi}{2}}(\omega)e^{\mathrm{i}(\omega\tau+\Delta\varphi)}\vec{e}_{\alpha+\frac{\pi}{2}}.
\end{equation}
Here, $\vec{e}_\alpha=[\cos(\alpha),\sin(\alpha)]^\mathrm{T}$ and $\vec{e}_{\alpha+\frac{\pi}{2}}=[-\sin(\alpha),\cos(\alpha)]^\mathrm{T}$ denote the rotated unit vectors, and $\tilde{E}^+_\mathrm{\alpha/\alpha+\frac{\pi}{2}}(\omega)$ are the retrieved complex-valued spectra.\\
To determine the parameters $\tau$ and $\Delta\varphi$, two additional fundamental spectra are recorded under projection angles of $\pm\frac{\pi}{4}$ with respect to $\alpha$ and compared to simulated spectra calculated as follows
\begin{align}
	I^\mathrm{sim}_{\alpha\pm\frac{\pi}{4}}(\omega;\Delta\varphi,\tau) =& \frac{1}{2}\left(\left|\tilde{E}^+_\mathrm{\alpha}(\omega)\right|^2+\left|\tilde{E}^+_\mathrm{\alpha+\frac{\pi}{2}}(\omega)\right|^2\right)\pm\left|\tilde{E}^+_\mathrm{\alpha}(\omega)\right|\left|\tilde{E}^+_\mathrm{\alpha+\frac{\pi}{2}}(\omega)\right|\notag\\
	&\times\cos\left\{\arg\left[\tilde{E}^+_\mathrm{\alpha}(\omega)\right]-\arg\left[\tilde{E}^+_\mathrm{\alpha+\frac{\pi}{2}}(\omega)\right]-\omega\tau-\Delta\varphi\right\}.
	\label{eq:DiagSpectra}
\end{align}
The parameters $\Delta\varphi$ and $\tau$ are determined by minimizing the root-mean-square error (RMSE), defined as
\begin{equation}
		\mathrm{RMSE}(\Delta\varphi,\tau)=\sqrt{\frac{1}{2N}\left[\sum\left(I^{\mathrm{meas}}_{\alpha+\frac{\pi}{4}} - I^{\mathrm{sim}}_{\alpha+\frac{\pi}{4}}\right)^2 + \sum\left(I^{\mathrm{meas}}_{\alpha-\frac{\pi}{4}} - I^{\mathrm{sim}}_{\alpha-\frac{\pi}{4}}\right)^2\right]},
		\label{eq:RMSE}
\end{equation}
where $N$ denotes the number of sample points for each of the spectra, and ${I}^\mathrm{meas}_\mathrm{\alpha\pm\frac{\pi}{4}}(\omega)$ are the measured power spectral densities. 
The optimal parameters $(\Delta\varphi,\tau)$ are found by selecting the absolute minimum of the RMSE on a discrete parameter grid of $501\times501$ points over the intervals $-\pi < \Delta\varphi < \pi$ and  $-10\,\mathrm{fs} < \tau < 10\,\mathrm{fs}$.
The grid is centered on a minimum identified beforehand over a larger $\tau$ interval with a width of up to \SI{200}{\femto\second}.
In general a single additional spectrum provides sufficient information to determine $\Delta\varphi$ and $\tau$. 
However, here we use two additional projections to perform the minimization on a larger data set and thereby obtain more robust estimates of the two parameters.
The retrieval determines the relative phase between the orthogonal field components but does not constrain the carrier-envelope phase.\\
The resulting temporal vector field is obtained via inverse Fourier transform, $\mathcal{F}^{-1}\left[\vec{\tilde{E}}^+(\omega)\right]=\vec{E}^+(t)$.
Projection onto the circular polarization basis $E_{L/R}^+(t)=\vec{E}^+(t)\cdot\vec{e}_\mathrm{L/R}$, with $\vec{e}_\mathrm{L/R}=\frac{1}{\sqrt{2}}(1,\mp\mathrm{i})^\mathrm{T}$ allows evaluation of the ellipticity according to \cite{Collett:2005} 
\begin{equation}
	\eta(t) = \frac{|E^+_\mathrm{L}(t)|^2-|E^+_\mathrm{R}(t)|^2}{|E^+_\mathrm{L}(t)|^2+|E^+_\mathrm{R}(t)|^2}.
	\label{eq:ellipticity}
\end{equation}
The ellipticity provides a convenient measure for comparing the retrieved and simulated polarization states. 
Simulated fields are obtained from
\begin{equation}
	\vec{\tilde{E}}^+_\mathrm{sim}(\omega) = \sqrt{I_\mathrm{meas}(\omega)}\boldsymbol{\mathsf{\Lambda}}_{\lambda/4}(\alpha_{\lambda/4})\left(e^{-\mathrm{i}\varphi^A(\omega)}\vec{e}_{-\pi/4}+e^{-\mathrm{i}\varphi^B(\omega)}\vec{e}_{+\pi/4}\right),
\end{equation}
where $I_\mathrm{meas}(\omega)$ is the measured PSD of the fundamental transform-limited laser pulse.\\ 
\section{Results}
\label{sec:results}
In this section, we present our results on the characterization of polarization-tailored femtosecond-laser pulses using the shaper-based d-scan technique.
First, we determine and compensate the residual spectral phase of the input pulse to compress the pulse close to its transform limit. 
The result is validated in section~\ref{subsec:compcharac} in a reference experiment where we apply linear spectral phases, in addition to the compensation phase, and employ an achromatic $\lambda/4$ wave plate to create a sequence of two compressed CRCP pulses.
Subsequently, in section~\ref{subsec:reconstruction}, we present the shaper-based generation and characterization of polarization-tailored pulses.
Finally, in view of application for attosecond photoelectron tomography, we apply our framework for the rotation of the PG pulse sequence by controlling the relative phase of the interfering circularly polarized pulses and demonstrate the generation of various polarization-tailored multi-pulse sequences in section~\ref{subsec:paravariation}.\\
\begin{figure*}[ht]
	\includegraphics{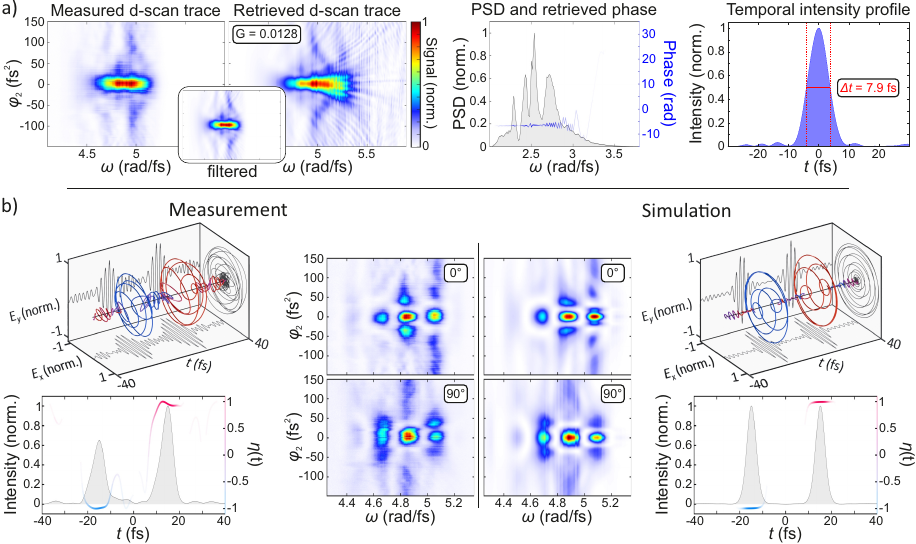}
	\caption{Pulse compression and reference experiment for verification of the shaper-based d-scan method. a) Measured and retrieved d-scan traces (inset shows retrieved trace weighted by retrieved scaling factor) along with the measured fundamental PSD and retrieved spectral phase as well as the temporal intensity profile. b) Measured and simulated CRCP laser pulse sequence with its corresponding d-scan traces and ellipticity, respectively.\label{fig2}}
\end{figure*}
\subsection{Pulse compression and reference experiment}
\label{subsec:compcharac}
D-scan traces are recorded by variation of the second-order dispersion parameter $\varphi_2$ in the range of $\varphi_2\in[\SI{-150}{\femto\second^2},\SI{150}{\femto\second^2}]$. Figure~\ref{fig2}a) presents the measured and retrieved d-scan traces, together with the fundamental PSD, the retrieved spectral phase, and the temporal intensity profile.\\
The retrieved d-scan trace reproduces the measured trace with a fitness of $G=0.0128$ (defined by the weighted RMSE between the measured and simulated traces, as in \cite{Miranda:2012:OE:688}), confirming the high fidelity of the spectral phase retrieval.
Due to absorption of UV light by the fiber collimator and the smaller conversion efficiency of the crystal, the expected weak second-harmonic signal at frequencies above \SI{5.2}{\radian\per\femto\second} is absent in the measured trace.
In contrast, the retrieved d-scan trace contains these contributions.
Nevertheless, accurate phase retrieval remains feasible, because the relevant phase information is also encoded in the remaining trace \cite{Miranda:2012:OE:18732,Miranda:2012:OE:688}.
Following a successful retrieval, the overall conversion efficiency is described by a frequency-dependent scaling factor (defined in \cite{Miranda:2012:OE:688}), which matches the simulated and measured traces. 
For clarity, all simulated and retrieved traces shown in the following are scaled using the retrieved scaling factors.
For comparison with the corresponding unfiltered trace a filtered trace is provided in the inset.\\
The retrieved spectral phase, shown in the central frame, is essentially flat, confirming the generation of a nearly transform-limited pulse, i.e., verifying the compensation of residual spectral phases introduced by the optical components in the beamline.
Regions of low spectral intensity, where the phase is not meaningful, are faded out for visualization purposes. 
The temporal intensity profile (last frame) obtained by inverse Fourier transformation exhibits a bell shape without significant pre- or post-pulses. 
The retrieved full width at half maximum (FWHM) is $\Delta t = \SI{7.9}{\femto\second}$.\\
Next, we investigate a (i) CRCP pulse sequence consisting of a pair of transform-limited pulses well-separated in time. 
Both pulses are circularly polarized but with opposite circularity, rendering it a prototype pulse for polarization-shaped laser pulses. 
This type of pulses was used in the literature, for example, for the creation of photoelectron vortices \cite{NgokoDjiokap:2015:PRL:113004, Pengel:2017:PRL:053003,Kerbstadt:2019:NC:658}. 
In this reference experiment, the pulse shaper is used for pulse compression and introduction of dispersion.
Figure~\ref{fig2}b) shows the measured and simulated d-scan traces for a CRCP pulse sequence generated using linear phase parameters of $\phi^{A}_1=\SI{15}{\femto\second}$ and $\phi^{B}_1=\SI{-15}{\femto\second}$ and the $\lambda/4$ wave plate aligned at an angle of $\alpha_{\lambda/4}=\SI{0}{\degree}$. By this means, the shaper generates a sequence of two orthogonal linearly polarized (OLP) pulse components separated by $\SI{30}{fs}$. Because this time-delay is much larger than the pulse duration, both pulses are well-separated in time. Subsequently, this OLP sequence is converted into a CRCP sequence by the $\lambda/4$ wave plate.
The d-scan traces were recorded for orthogonal polarization projections at $\SI{0}{\degree}$ and $\SI{90}{\degree}$ by rotating the polarizer and the BBO crystal. The reconstructed vectorial electric field of the CRCP sequence is shown in the 3D boxes. The corresponding ellipticities, shown as faded curves in the bottom frames together with the total pulse intensity (gray-shaded background), were calculated using Eq.~\eqref{eq:ellipticity}. 
The measured traces are consistent with the simulations based on the independently measured fundamental spectrum of the input pulse (indicated by the faded spectrometer fiber in Fig.~\ref{fig1}).
The spectral phases of the orthogonal polarization components are retrieved individually.
The full vectorial electric field is reconstructed by determining the relative constant and linear spectral phase offsets from additional measurements at $\pm\SI{45}{\degree}$, following the procedure in \cite{DiazRivas:2024:JPP:015003} and described above.
The vectorial field reconstructed from the measured traces is shown in the upper left frame.
It is in excellent agreement with the simulated field shown in the upper right frame.
For a quantitative evaluation, the ellipticities are depicted separately in the bottom frames.
Since ellipticity is only meaningful in time intervals of non-vanishing intensity, the temporal intensity profile (gray-shaded background) is used to adapt the visibility in low-intensity regions, for visualization purposes.
We observe a high-quality retrieval of the CRCP pulse sequence with an almost perfect ellipticity of $|\eta|\approx1$, i.e., $|\eta|>0.92$ within the FWHM of the individual pulses.
This example served to validate our shaper-based d-scan method for the retrieval of the vectorial character of polarization-tailored laser pulses.
\begin{figure*}[htpb]
	\includegraphics[width=\linewidth]{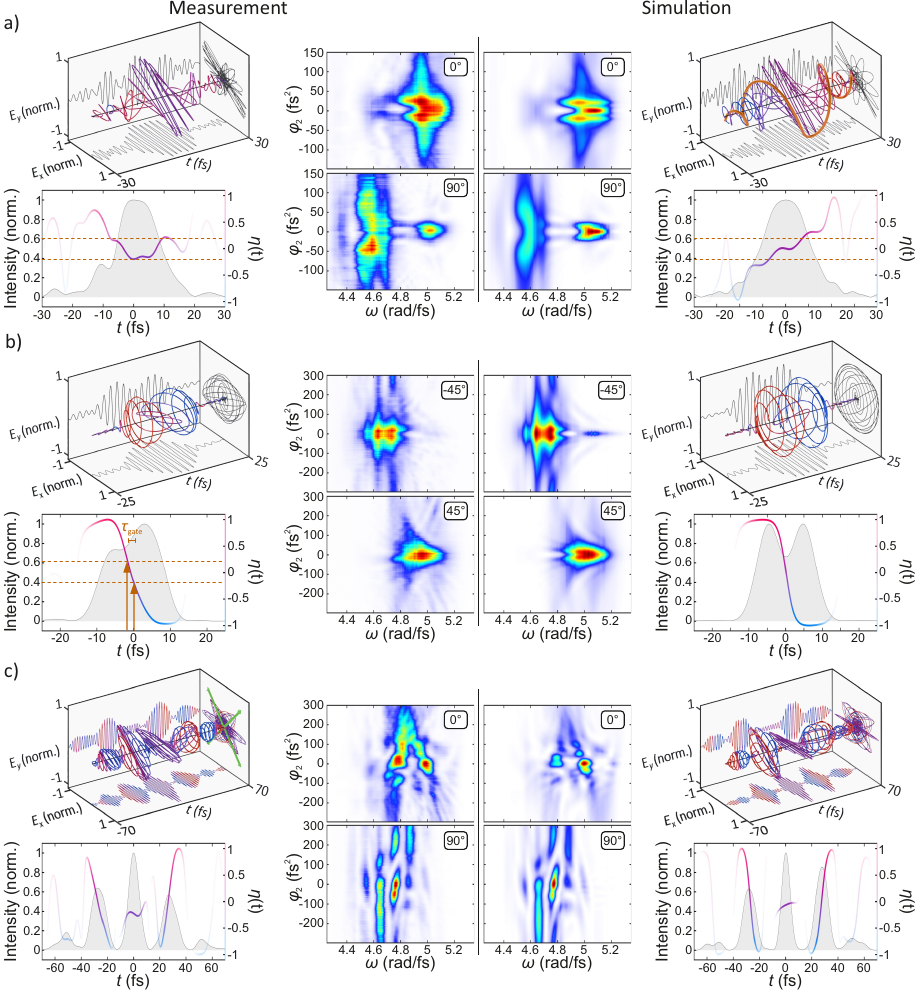}
	\caption{Measured (left) and simulated (right) d-scan traces together with retrieved laser electric field and corresponding ellipticity. a) OC-CRCP pulse, b) PG pulse, c) Multi-pulse sequence.\label{fig3}}
\end{figure*}
\subsection{Reconstruction of polarization-tailored laser fields}
\label{subsec:reconstruction}
Here, we present our results on the generation and characterization of the following physically motivated polarization-tailored pulses: 
(ii) an OC-CRCP pulse as used in \cite{Strandquist:2022:PRA:043110,Koehnke:2024:PRA:053109} for the creation of reversible photoelectron spirals and to drive non-perturbative adiabatic dynamics \cite{Koehnke:2025:PRA:023104}, (iii) a PG pulse originally introduced in \cite{Corkum:1994:OL:1870} and experimentally demonstrated in \cite{Sansone:2006:Science:443} for the creation of isolated attosecond pulses via high-harmonics generation (HHG) and (iv) multi-pulse sequences generated via sinusoidal spectral phase-modulation, as routinely applied in coherent control of atoms and molecules \cite{Meshulach:1998:Nature:239,Dudovich:2005:PRL:083002,Bayer:2016:ACP:235,Herek:2002:Nature:533,Hauer:2006:JCP:061101}, demonstrating the versatility of the shaper-based approach for more complex pulse shapes. 
The results are shown in Fig.~\ref{fig3}a)-c), respectively, and are discussed individually in the following paragraphs.
\paragraph{OC-CRCP pulse} 
We start with the OC-CRCP pulse generated by superposition of two circularly polarized pulse components with opposite circularity and opposite chirp. The time-delay is set to zero so that the two pulse components overlap perfectly in time. 
This type of pulse was proposed in \cite{Strandquist:2022:PRA:043110} for the creation of reversible electron spirals, which were experimentally demonstrated in \cite{Koehnke:2024:PRA:053109}. 
In the strong-field regime, such pulses were employed to control adiabatic dynamics in a V-type three-level system \cite{Koehnke:2025:PRA:023104}.
OC-CRCP pulses exhibit several characteristic features as discussed in \cite{Koehnke:2025:PRA:023104} which render this pulse class a suitable benchmark for the retrieval of polarization-tailored fields.
For a symmetric fundamental spectrum, the OC-CRCP pulse is linearly polarized at every instant in time, while the polarization direction continuously rotates around the propagation axis. The rotation decelerates towards $t=0$, reverses its direction, and accelerates again for $t>0$, reflecting the time-reversal symmetry of the applied quadratic phases.
The instantaneous frequency remains constant, as the spectral components are delayed symmetrically in time with opposite chirps, resulting in a superposition oscillating at the central frequency. \\
The OC-CRCP pulse is generated by setting $\phi^{A}_1=\phi^{B}_1=0$ and using quadratic phase parameters of $\phi_2^{A}=\SI{20}{\femto\second^2}$ and $\phi_2^{B}=\SI{-20}{\femto\second^2}$, in combination with the $\lambda/4$ wave plate aligned at $\alpha_{\lambda/4} = \SI{0}{\degree}$.
The experimental and simulation results are presented in Fig.~\ref{fig3}a). The measured and calculated d-scan traces are in very good agreement.
The reconstructed vectorial field indicates the expected linear polarization in the temporal window of maximum field intensity around $t=0$. This observation is confirmed by the ellipticity being close to zero (violet) in this region.
Quantitatively the ellipticity satisfies $|\eta|<0.2$ during the central part of the pulse, as indicated by the horizontal orange dashed lines. This value is a commonly used threshold for an effectively linear polarization in HHG \cite{Oron:2006:PRA:063816, Moeller:2012:PRA:011401,Ivanov:1995:PRL:2933}.
The small residual ellipticity arises mainly from the asymmetry of the fundamental spectrum.
Applying a quadratic spectral phase to an asymmetric spectrum leads to a chirped pulse with temporally asymmetric envelope. Because inversion of the chirp reverses the pulse -- and hence the asymmetry -- in time, the superposition of the two oppositely chirped left- and right-handed circularly polarized components do not yield perfect linear polarization. 
We confirmed this argument numerically by using symmetric fundamental spectra in our simulations.
Nevertheless, the characteristic rotation of the polarization direction, highlighted by the temporal envelope (solid orange line) in the simulated field, is clearly observed in both the measured and the simulated field.\\
\paragraph{Polarization gate} 
Next, we apply the shaper-based d-scan technique to characterize a PG pulse sequence.
To this end, the chirp is set to zero, $\phi^{A}_2=\phi^{B}_2=0$, and linear phase parameters of $\phi^{A}_1=\SI{4.75}{\femto\second}$ and $\phi^{B}_1=\SI{-4.75}{fs}$ are used. The $\lambda/4$ wave plate is aligned at $\alpha_{\lambda/4}=\SI{90}{\degree}$. Due to the linear spectral phases, the left- and right-handed circularly polarized pulse components are delayed by $\SI{9.5}{fs}$. This time-delay is comparable to the pulse duration. Therefore, both pulses partially overlap in time forming a PG in the short time window around $t=0$ where both amplitudes are approximately equal. 
Since the PG pulse is oriented horizontally, the d-scans are performed for $\alpha=-\pi/4$, i.e., for projections under $\SI{\pm45}{\degree}$, ensuring similar signal amplitudes for a robust retrieval. \\
The results are presented in Fig.~\ref{fig3}b). The formation of the PG is clearly visible in both the reconstructed and simulated vectorial fields.
A quantitative evaluation is obtained from the temporal evolution of the ellipticity.
We observe a significantly smaller time window of linear polarization than the bandwidth limited pulse duration.
Specifically, the interval for which $|\eta|<0.2$ amounts to $\tau_\mathrm{gate}=\SI{1.63}{\femto\second}$, corresponding to about one-fifth of the transform-limited pulse duration and two thirds of an optical cycle, for the measured PG pulse, highlighting its use for HHG.
In Sec.~\ref{subsec:paravariation}, we show that the shaper-based generation of the PG pulse sequence enables precise control over orientation of the linear polarization of the PG pulse, highlighting the versatility of the approach and enable attosecond photoelectron tomography.
\paragraph{Multi-pulse sequence} 
Finally, we investigate a polarization-tailored multi-pulse sequence generated by applying individual sinusoidal spectral phase functions to the orthogonal polarization components. 
The sine amplitudes are set equally to $A_\mathrm{sin}^{A/B}=1.44$. 
The remaining sine parameters are set to $\tau_\mathrm{sin}^{A}=\SI{30}{\femto\second}$, $\tau_\mathrm{sin}^{B}=\SI{25}{\femto\second}$, $\phi_\mathrm{sin}^{A}=0$ and $\phi_\mathrm{sin}^{B}=\frac{\pi}{2}$. 
As a result, two OLP pulse sequences with slightly different temporal separations ($\tau_\mathrm{sin}^A\neq\tau_\mathrm{sin}^B$) are generated. 
The sine amplitude is chosen such that the three central subpulses of each pulse sequence have equal peak-amplitude, according to the relation $|J_{\pm1}(A_\mathrm{sin})|\approx|J_0(A_\mathrm{sin})|$ \cite{Wollenhaupt:2006:PRA:063409} with $J_n$ being the Bessel function of the first kind and $n$-th order.
Again, the $\lambda/4$ wave plate aligned at an angle of $\alpha_{\lambda/4}=\SI{0}{\degree}$ converts the orthogonal linearly polarized pulse components into left- and right-handed circular components. 
Due to the slightly different pulse-to-pulse separations in the sequences of the two polarization components, their superposition results in a vectorial field forming multiple PGs and a central subpulse that is linearly polarized. \\
The results are presented in Fig.~\ref{fig3}c).
The reconstructed vectorial field clearly reveals multiple PGs. 
Their orientation is best discernible in the projection along the propagation direction and is indicated by green arrows as a guide to the eye; in the case investigated here, the gates are aligned orthogonally.
Such sequences of PG pulses offer promising opportunities for attosecond science.
In particular, they enable the generation of attosecond pulses with well-defined and precisely controllable time-delays, which can be exploited in attosecond double- or multi-slit experiments \cite{Kaneyasu:2023:SR:6142}. 
Moreover, by adjusting the relative phase of the applied sinusoidal phase masks, the orientation of the individual PGs can be controlled. 
This provides an additional degree of control for attosecond photoionization experiments, in which interference between electron wave packets with different angular distributions -- generated by time-delayed attosecond pulses of distinct polarization -- can be investigated.
\subsection{Polarization-gate rotation and polarization control of multi-pulse sequences}
\label{subsec:paravariation}
In this section, we demonstrate the versatility of the programmable pulse-shaping and characterization scheme through controlled rotation of the PG and the generation of tailored multi-pulse sequences.
To this end, we vary the relative phase between the two CRCP subpulses forming the PG and investigate parameter variation for sinusoidal phase modulations and a discrete periodic phase function.\\
\begin{figure*}[htpb]
	\includegraphics[width=\linewidth]{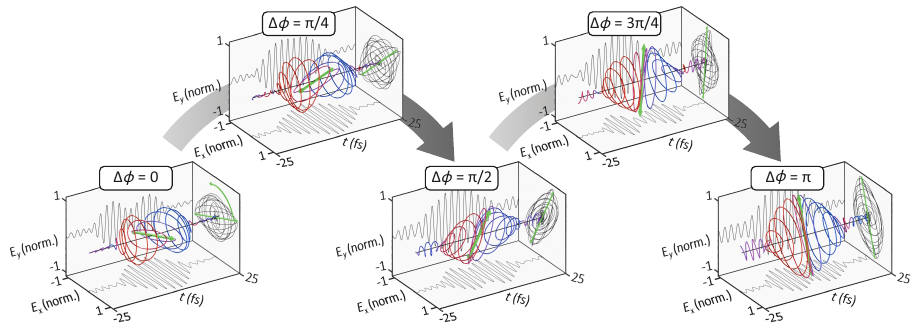}
	\caption{Controlling the rotation of the PG by variation of the relative phase in steps of $\Delta\phi=\pi/4$ leading to rotations of $\Delta\phi/2$. Expected orientations are indicated by greens arrows, respectively.\label{fig4}}
\end{figure*}
Figure~\ref{fig4} shows the reconstructed PG pulses generated using the fixed linear phase parameters $\phi_1^{A}=\SI{4.75}{\femto\second}$ and $\phi_1^{B}=\SI{-4.75}{\femto\second}$, while varying the relative phase $\Delta\phi=\phi_0^A-\phi_0^B$ over the values $\{0,\frac{\pi}{4},\frac{\pi}{2},\frac{3\pi}{4},\pi\}$.
The orientation of the PG is given by \cite{Collett:2005} 
\begin{equation}
	\alpha_\mathrm{gate} = \frac{1}{2}\arctan\left(\frac{S_2}{S_1}\right)+n\frac{\pi}{2}=\frac{\Delta\phi}{2}+n\frac{\pi}{2},
\end{equation}
where $S_1\propto\cos(\Delta\phi)$ and $S_2\propto\sin(\Delta\phi)$ are the relevant Stokes parameters, with $n=1$ for $S_1<0$ and $n=0$ otherwise.
The gate rotates by $\Delta\phi/2$, such that, for example, the gates for $\Delta\phi=0$ and $\Delta\phi=\pi$ are orthogonal. 
The targeted gate orientation is indicated by green arrows in the projection along the propagation direction.\\
The slight offset from horizontal orientation at $\Delta\phi=0$ results from a slight mismatch between $\omega_\mathrm{mod}$ and the spectral central frequency (see Eq.~\eqref{eq:Phaseparameterization}).
These results demonstrate direct control over the orientation of the PGs and highlight the potential of shaper-generated tailored driver pulses for attosecond ionization with polarization-controlled HHG.
In this way PG control enables a route toward attosecond photoelectron tomography analogously to femtosecond photoelectron tomography \cite{Wollenhaupt:2009:APB:647}.
In general, HHG is highly sensitive to focusing and phase-matching conditions \cite{Constant:1999:PRL:1668,Balcou:1997:PRA:3204}. 
Therefore, implementing PG control without moving optical components is very desirable.\\
In addition, we investigate parameter variations of independent sinusoidal and periodic square-wave spectral phase modulations applied to the two orthogonally polarized field components.
The results are presented in Fig.~\ref{fig5}.
In Fig.~\ref{fig5}a), sine parameters of $A_\mathrm{sin}^{A/B}=1.44$, $\tau_\mathrm{sin}^{A/B}=\SI{30}{\femto\second}$ and $\phi_\mathrm{sin}^{B}=\frac{\pi}{2}$ were used and the $\lambda/4$ wave plate was aligned at $\alpha_{\lambda/4}=\SI{90}{\degree}$.
By this means, a sequence of linearly polarized subpulses with varying orientations is generated. 
To guide the eye, time windows where the ellipticity is $|\eta|<0.2$ are again highlighted by dotted orange lines in the lower panels.
The imperfections in linear polarization are attributed to the asymmetric spectrum and small side lobes of the subpulses, which are also observed in the simulated pulse sequence. \\
\begin{figure*}[htpb]
	\includegraphics[width=\linewidth]{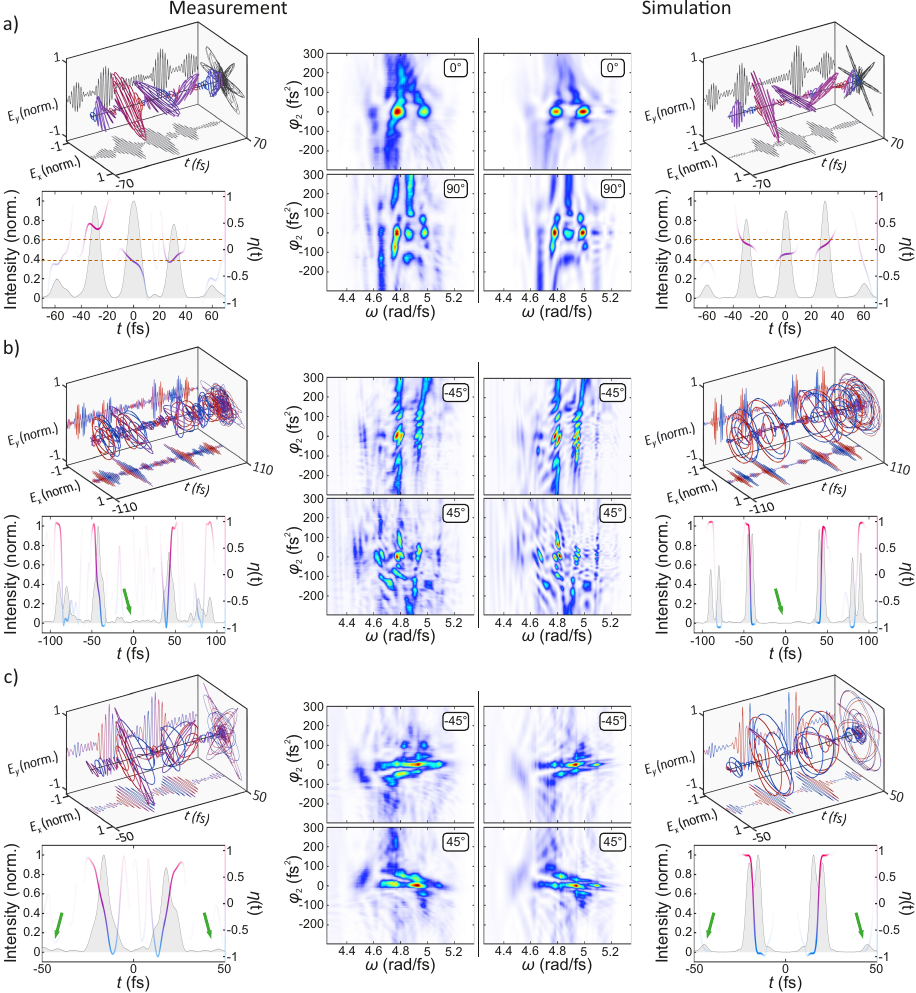}
	\caption{Variation of sinusoidal and square-wave phase modulations generating polarization-tailored multi-pulse sequences. a) Sequence of linearly polarized pulses, b) Sequence of PG pulses with varying orientation, c) Double-pulse of orthogonal oriented PG pulses generated via square-wave spectral phase modulation. \label{fig5}}
\end{figure*}
In Fig.~\ref{fig5}b), sinusoidal phase modulations with $A_\mathrm{sin}^{A/B}=2.41$ are used to suppress the central subpulse (highlighted by green arrows), since $J_0(2.41)\approx0$.
Together with slightly different modulation periods, $\tau_\mathrm{sin}^{A}=\SI{45}{\femto\second}$ and $\tau_\mathrm{sin}^{B}=\SI{40}{\femto\second}$ and $\alpha_{\lambda/4}=\SI{0}{\degree}$, this choice generates multiple PGs.
Despite the highly structured d-scan traces arising from the increased subpulse separations, the measured and simulated traces, reconstructed fields, and retrieved ellipticities are again in excellent agreement, demonstrating robustness and fidelity of the retrieval even for highly structured polarization-tailored laser pulses.\\
Finally, Fig.~\ref{fig5}c) shows the results obtained using square spectral phase modulation,
\begin{equation}
\varphi_\mathrm{mod}(\omega) = A^{A/B}_\mathrm{sin}\sigma\left[\tau^{A/B}_\mathrm{sin}(\omega-\omega_\mathrm{mod})+\phi^{A/B}_\mathrm{sin}\right],
\label{eq:squaremodulation}
\end{equation}
where $\sigma$ denotes the sign function.
This binary phase modulation serves two purposes: First, it generates a double-pulse sequence with negligible satellite pulse contributions \cite{Pestov:2009:OE:14351} (see green arrows for small side lobes). 
Second, it is a prototype for discrete phase modulation functions. 
On this  example we demonstrate the shaper-based polarization manipulation along with the applicability of the retrieval even for discontinuous spectral phase functions.
Using $\alpha_{\lambda/4}=\SI{0}{\degree}$, $A_\mathrm{sin}^{A/B}=\frac{\pi}{2}$, $\tau_\mathrm{sin}^{A}=\SI{20}{\femto\second}$, $\tau_\mathrm{sin}^{B}=\SI{15}{\femto\second}$ and $\phi_\mathrm{sin}^{B}=\frac{\pi}{2}$, the measured and simulated traces are again in good agreement.
Particularly striking is the reproduced mirror symmetry of the $\SI{\pm45}{\degree}$ projections, as well as the reconstruction of the two desired PGs.\\
Overall, the results presented in this section demonstrate that the shaper-based d-scan framework enables compression, generation, and characterization of highly structured vectorial femtosecond-laser fields, including controlled PG rotation, polarization-tailored multi-pulse sequences, and discontinuous spectral phase modulation, thus extending the toolbox for ultrafast optics and attosecond science.
\section{Summary and Conclusion}
\label{sec:conclusion}
In this work, we combined ultrafast polarization pulse-shaping with d-scan-based pulse characterization.
Building on the approach introduced in \cite{DiazRivas:2024:JPP:015003}, we established a framework for the simultaneous generation and characterization of polarization-tailored femtosecond-laser pulses.
A $4f$ polarization pulse shaper was used both to control the spectral phase of two orthogonal polarization components and to introduce the second-order dispersion required to record the d-scan trace.
The vectorial laser fields were reconstructed from the measurement of two orthogonal projections, while the remaining constant and linear relative phases were retrieved by matching them to additional spectra measured at two further projection angles.\\
To validate the performance of the framework, we demonstrated the retrieval of different physically motivated polarization-tailored pulses: (i) CRCP pulse sequences, used for the generation of photoelectron vortices \cite{Pengel:2017:PRL:053003,Kerbstadt:2019:NC:658}; (ii) OC-CRCP pulses, employed for reversible photoelectron spirals and strong-field adiabatic dynamics \cite{Strandquist:2022:PRA:043110,Koehnke:2024:PRA:053109,Koehnke:2025:PRA:023104}; (iii) PG pulse sequences for the generation of isolated attosecond pulses in HHG \cite{Corkum:1994:OL:1870,Sansone:2006:Science:443}; and (iv) multi-pulse sequences generated via sinusoidal spectral phase modulation, routinely applied in coherent control of atoms and molecules \cite{Meshulach:1998:Nature:239,Dudovich:2005:PRL:083002,Bayer:2016:ACP:235,Herek:2002:Nature:533,Hauer:2006:JCP:061101}.
All pulse types were consistently reconstructed, and their characteristic features were accurately reproduced.\\
Furthermore, we demonstrated control over the orientation of a PG sequence via the relative phase between CRCP components, enabling attosecond photoelectron tomography.
The versatility of the approach is further illustrated by the generation and characterization of multi-pulse sequences with circular and linear polarization, as well as sequences of PG pulses, with controlled relative orientations, respectively.\\
The presented framework combines the flexibility of programmable pulse shaping with the robustness of the d-scan characterization method, and provides a single experimental platform for the generation and retrieval of advanced vectorial femtosecond laser fields.
Our approach provides a versatile tool for experiments requiring ultrashort pulses with controlled polarization states and thus will be of interest for applications in ultrafast optics and coherent control of light-matter interactions. 

\ack{
We acknowledge helpful discussions with Jan Vogelsang regarding the d-scan retrieval algorithm.
}

\funding{
This work was funded by the Wissenschaftsraum \textit{Elektronen-Licht-Kontrolle} (elLiKo) funded by the Niedersächsische Ministerium für Wissenschaft und Kultur and the  Deutsche Forschungsgemeinschaft via the project 467215508.
}

\roles{
D Köhnke \orcid{0000-0003-1151-8625}\\
\noindent Conceptualization (lead), Data Curation (supporting), Formal Analysis (lead), Investigation (supporting), Methodology (lead), Project Administration (lead), Software (supporting), Supervision (lead), Validation (lead), Visualization (lead), Writing -- original draft, Writing -- review \& editing (lead)\\[0.5em]
\noindent J-E Havekost \orcid{0009-0003-2040-2920}\\
\noindent Formal Analysis (supporting), Software (lead), Writing -- review \& editing (supporting)\\[0.5em]
\noindent N R Rother \orcid{0009-0003-8469-5558}\\
\noindent Data Curation (lead), Investigation (lead), Methodology (supporting), Software (supporting), Validation (supporting), Visualization (supporting), Writing -- review \& editing (supporting)\\[0.5em]
\noindent C Kuntze \orcid{0009-0009-8452-4998}\\
\noindent  Investigation (supporting), Methodology (supporting), Resources (supporting), Validation (supporting), Writing -- review \& editing (supporting)\\[0.5em]
\noindent L Englert \orcid{0000-0002-2275-9136}\\
\noindent Funding Acquisition (supporting), Software (supporting), Supervision (supporting), Writing -- review \& editing (supporting)\\[0.5em]
\noindent T Bayer \orcid{0009-0005-2280-5672}\\
\noindent Conceptualization (supporting), Funding Acquisition (supporting), Supervision (supporting), Writing -- review \& editing (supporting)\\[0.5em]
\noindent M Wollenhaupt \orcid{0000-0002-0839-1494}\\
\noindent Conceptualization (supporting), Funding Acquisition (lead), Resources (lead), Supervision (supporting), Writing -- review \& editing (supporting)
}

\data{The datasets used during the current study are available from the corresponding author on reasonable request.}

\bibliographystyle{ieeetr}
\bibliography{ultra_db}
\end{document}